\documentclass[reprint,aps,pra,showpacs,floatfix]{revtex4-1}

\usepackage{graphicx, amsmath,amssymb,psfrag,empheq,hyperref,natbib}
\usepackage{dcolumn}
\usepackage{bm}
\usepackage{epsfig}
\begin{document}

\title{Exact Solutions to Non-Linear Symmetron  Theory II: \\
One and Two Mirror Systems}

\author{Mario Pitschmann}
\email{mario.pitschmann@tuwien.ac.at}
\affiliation{Atominstitut, Technische Universit\"at Wien, Stadionallee 2, A-1020 Wien, Austria}


\begin{abstract}
We derive the exact analytical solutions to the symmetron field theory equations in the presence of a one or two mirror system in the case of a spontaneously broken phase in vacuum as well as in matter. This complements a similar analysis performed in a previous article \cite{Brax:2017hna}, in which the symmetron is in the spontaneously broken phase in vacuum but in the symmetric phase in matter.  Here again, the one dimensional equations of motion are integrated exactly for both systems and their solutions are expressed in terms of Jacobi elliptic functions. In the case of two parallel mirrors the equations of motion provide also in this case a discrete set of solutions with increasing number of nodes and energies. The solutions obtained herein can be applied to \textit{q}BOUNCE experiments, neutron interferometry and to the calculation of the symmetron field induced "Casimir force" in the \textsc{Cannex} experiment and allow to extend the investigation to hitherto unavailable regions in symmetron parameter space.     
\end{abstract}


\maketitle


\section{Introduction}

Cosmological observations reveal that our universe is currently expanding at an accelerated rate. The theoretical framework describing the universe on cosmological scales is general relativity. Since cosmological observations rely on general relativity for the interpretation of  the experimental results it appears natural that general relativity might need to be modified to account for the current accelerated expansion of the universe. While a modification for short distances is indeed easily possible, the modification of the theory for large distance scales is very intricate and would violate some of the theory's fundamental assumptions. It appears more natural to consider the existence of additional new hypothetical scalar fields, which couple to gravity and can account for dark energy (see \cite{Joyce:2014kja} for further details). The existence of hypothetical new scalar field degrees of freedom is strongly motivated beyond their cosmological application. The presence of a new scalar typically induces also new interactions, so-called fifth forces. On the other hand, experiments at solar distance scales and below give no evidence for fifth-forces and as such provide strong constraints on these
\cite{Will:2014kxa}. Naturally, is is assumed that some kind of "screening mechanism" is at work, which suppresses the scalar and/or its interaction with matter in experimentally accessible regions 
of comparably high mass density. At cosmological scales the scalar prevails and contributes to the accelerated expansion. 

Among the manifold screening mechanisms \cite{Joyce:2014kja} are chameleon \cite{Khoury:2003rn,Khoury:2003aq,Brax:2004qh}, Damour-Polyakov \cite{Damour:1994zq}, K-mouflage \cite{Babichev:2009ee,Brax:2012jr,Brax:2014wla} and Vainshtein \cite{Vainshtein:1972sx} mechanisms. Among the models employing the Damour-Polyakov mechanism, which relies on a weakened coupling to matter in high density regions, is the symmetron model \cite{Hinterbichler:2010es, Hinterbichler:2011ca} (for earlier work see \cite{Pietroni:2005pv, Olive:2007aj}). 
In this model the coupling of the scalar to matter is proportional to the vacuum expectation value (VEV) of the field. The effective potential resembles the Higgs mechanism for a real rather than a complex field. In regions of low mass density spontaneous symmetry breaking occurs resulting in a non-vanishing VEV. In this case the field prevails and when its mass is small ($\sim10^3 H_0$) it could have cosmological implications \cite{Hinterbichler:2011ca} especially on the growth of perturbations and large-scale structure \cite{Clampitt:2011mx,Taddei:2013bsk}. 
In contrast in regions of high mass density the symmetry is restored and the VEV, as well as the scalar coupling to matter vanish, rendering the symmetron invisible to experimental observation (see e.g. \cite{Brax:2017hna} for further details on symmetrons).  

In \cite{Burrage:2016rkv,Brax:2016wjk}, atomic interferometry was investigated theoretically to constrain symmetrons. Bounds on symmetrons have been obtained by Jaffe \textit{et al.} \cite{Jaffe:2016fsh}. Gravity resonance spectroscopy (GRS) \cite{Abele:2009dw, Jenke:2011zz} has been employed and provided limits on symmetrons in \cite{Cronenberg:2018qxf}. In the corresponding theoretical analysis \cite{Brax:2017hna} exact analytical solutions have been found for an idealized 1-dimensional setup of a single mirror covering an infinite half-space or two parallel mirrors of finite separation each covering an infinite half-space. 

Supposedly in any experimental analysis performed so far, the symmetron was considered exclusively to be in its symmetric phase inside the matter regions of the experimental setup. This limitation restricted the experimentally obtained limits of GRS \cite{Cronenberg:2018qxf}. Following earlier investigations we do not constrain the analysis to regions in parameter space with respect to their cosmological relevance. In this article the symmetron is analyzed in the presence of a one-or two-mirror system with the symmetron in its spontaneously broken phase throughout, i.e. in vacuum and also in matter. It leads to new solutions, which can again be expressed in terms of Jacobi elliptic functions. Thereby, a new region of the symmetron parameter space can be experimentally probed. The solutions obtained here are valid only for the symmetron in its spontaneously broken phase in vacuum and matter. Consequently, this analysis is complementary to the previously performed one \cite{Brax:2017hna}. The solutions obtained in the latter together with those obtained herein provide a comprehensive analysis to obtain complete experimental bounds. 

In section \ref{sec:1} we will recall some background information on symmetrons, which will provide the relevant definitions for the field theory analysis. In section \ref{sec:2} the solutions for the one mirror case will be derived, while in sections \ref{sec:3} and \ref{sec:4} the symmetric and anti-symmetric two mirror solutions will be obtained. As an illustration, in section \ref{sec:5} a particular case study is carried out for an arbitrary choice of parameters and the complete spectrum of solutions for these parameters is derived. Section \ref{sec:6} provides relevant information on the \textit{q}BOUNCE experiment, where the corresponding symmetron induced resonance frequency shift for the case of a single mirror is obtained. In section \ref{sec:7} relevant information on the \textsc{Cannex} experiment is provided and the symmetron induced pressure derived. A conclusion in section \ref{sec:8} will be followed by two Appendices providing additional technical details on the symmetron induced force on a point particle in \ref{sec:A} and the screening of a neutron in \ref{sec:B}.

\section{Background}\label{sec:1}

Following \cite{Joyce:2014kja} (see also \cite{Brax:2017hna}), the symmetron effective potential is given by 
\begin{align}
\label{eq:VEFF}
   V_\text{eff}(\phi) &= V(\phi) + A(\phi)\,\rho \nonumber\\
   &= \frac{1}{2}\left(\frac{\rho}{M^2} - \mu^2\right)\phi^2 + \frac{\lambda}{4}\,\phi^4\>,
\end{align}
with the inverse coupling of dimension mass $M$ to the environmental mass-density $\rho$, 
a parameter $\mu$ of dimension mass and the dimensionless self-interaction coupling $\lambda$
(an additional term proportional to $\rho$, which will not affect the equations of motion has been neglected).
The Weyl-rescaling factor for the metric is for symmetrons defined as 
\begin{align}\label{eq:WRF}
   A(\phi) = 1 + \frac{\phi^2}{2M^2} + \mathcal O(\phi^4/M^4)\>.
\end{align}
For $\rho \geq M^2\mu^2$  the symmetron is in the "symmetric phase".  In the ”broken symmetry phase”, which we consider exclusively in this article, 
we have instead $\rho < M^2\mu^2$. We introduce the notation
\begin{align}\label{mui}
   \mu_i^2 = \mu^2 - \frac{\rho_i}{M^2}\>,
\end{align}
where in the ”broken symmetry phase” $\mu_i^2  > 0$.
The 1-dimensional Hamiltonian reads \cite{Brax:2017hna}
\begin{align}\label{Ham}
   \mathcal H = \frac{1}{2}\left(\frac{d\phi}{dz}\right)^2 - \frac{\mu_i^2}{2}\,\phi^2 + \frac{\lambda}{4}\,\phi^4\>.
\end{align}
The minimum value is given by $V_{\text{eff},\phi}(\phi)=0$ and reads $\pm\phi_i$ where (without loss of generality we employ $\mu_i\geq0$)
\begin{align}
\label{eq:phimin}
   \phi_i = \frac{\mu_i}{\sqrt{\lambda}}\>.
\end{align}
One should note, that while the minimum does not vanish inside matter for $\rho_i < M^2\mu^2$, due to Eqs.~(\ref{mui}) and (\ref{eq:phimin}), it takes a lower value in matter than in vacuum. The equations of motion for static field solutions we are searching for are given by
\begin{align}
   \frac{d^2\phi}{dz^2} = V_{\text{eff},\phi}(\phi)\>.
\end{align}
After multiplication by $\displaystyle\frac{d\phi}{dz}$ these can be integrated to
\begin{align}\label{1LD1}
   \frac{1}{2}\left(\frac{d\phi}{dz}\right)^2 - \frac{1}{2}\left(\frac{d\phi}{dz}\right)^2\bigg|_{z=z_0} = V_\text{eff}(\phi) - V_\text{eff}(\phi)\big|_{z=z_0}\>.
\end{align}
Given the value of the field and its first derivative at some point $z_0$ determines the solutions to the field equations completely. 

For later use we summarize the most relevant relations concerning Jacobi elliptic functions.
The Elliptic Integral of the first kind is given by
\begin{align}
   F(\phi,\ell) = \int_0^{\sin\phi}\frac{dt}{\sqrt{(1 - t^2)(1 - \ell^2t^2)}}\>.
\end{align}
Its relation to those Jacobi elliptic functions relevant for our analysis is as follows
\begin{align}
   \textrm{sn}(u,\ell) &= \sin\big(F^{-1}(u,\ell)\big) = - \textrm{sn}(-u,\ell)\>, \nonumber\\
   \textrm{cn}(u,\ell) &= \cos\big(F^{-1}(u,\ell)\big) = + \textrm{cn}(-u,\ell)\>, \nonumber\\
   \textrm{dn}(u,\ell) &= \sqrt{1 - \ell^2\sin^2\!\big(F^{-1}(u,\ell)\big)} = + \textrm{dn}(-u,\ell)\>.
\end{align}
Furthermore, the following useful relation holds 
\begin{align}
   \textrm{sn}\Big(u + F\Big(\frac{\pi}{2},\ell\Big),\ell\Big) = \frac{\textrm{cn}(u,\ell)}{\textrm{dn}(u,\ell)} = \textrm{cd}(u,\ell)\>.
\end{align}
Hyperbolic functions may be expressed in terms of Jacobi elliptic functions as well, i.e.
\begin{align}
   \tanh u &= \textrm{sn}(u,1)\>, \nonumber\\
   \coth u &= \textrm{ns}(u,1)\>.
\end{align}

\section{1 Mirror}\label{sec:2}

Analogously to \cite{Brax:2017hna}, we treat the case of a single mirror filling the infinite half-space $z \leq 0$ in this section.

\subsection{Vacuum \& Mirror Solutions}

In \cite{Brax:2017hna}, the solution for vacuum of low density $\rho_V$ in the infinite half-space $z \geq 0$ was found to be
\begin{align}
   \phi(z) = \phi_V\frac{k_V}{|k_V|}\tanh\!\Big(\frac{\mu_V z}{\sqrt2} + \tanh^{-1}|k_V|\Big)\>,
\end{align}
where $k_V:= \phi_0/\phi_V$ is the ratio between the value of $\phi$ taken for $z=0$, i.e. $\phi_0$, and the minimum value in vacuum $\phi_V = \mu_V/\sqrt\lambda$.

Next, we consider the case of low density within the mirror where $\rho_M < M^2\mu^2$ and 
search for a solution that asymptotically for $z \to -\infty$ goes as $\phi(z) \to \pm\phi_M = \pm\mu_M/\sqrt\lambda$ 
implying $\displaystyle\frac{d\phi}{dz} \to 0$. Hence we find
\begin{align}
   \frac{1}{2}\left(\frac{d\phi}{dz}\right)^2 = V_\text{eff}(\phi) - V_\text{eff}(\phi_M)\>,
\end{align}
leading to
\begin{align}
   \int_{\phi_0}^{\phi(z)}\frac{d\phi}{\sqrt{-\mu_M^2\,\big(\phi^2 - \phi_M^2\big) + \lambda/2\,\big(\phi^4 - \phi_M^4\big)}} = \frac{k_M}{|k_M|}\,z\>,
\end{align}
where $k_M:= \phi_0/\phi_M$ is the ratio between the value of $\phi$ taken for $z=0$ and the minimum value inside the mirror $\phi_M$. Furthermore, we obtain 
\begin{align}
   \frac{k_M}{|k_M|}\,z &= \frac{1}{\mu_M}\int_{k_M}^y\frac{dy'}{\sqrt{\displaystyle1 - y'^2 + 1/2\,\big(y'^4 - 1\big)}}  \nonumber\\
   &= -\frac{\sqrt2}{\mu_M}\int_{k_M}^y\frac{dy'}{\displaystyle1 - y'^2}  \nonumber\\
   &= -\frac{\sqrt2}{\mu_M}\,\big(\coth^{-1}y - \coth^{-1}k_M\big)\>,
\end{align}
where $y:= \phi(z)/\phi_M$. In the last line we made use of the fact that $|y|>1$.
Inverting the relation straightforwardly leads to
\begin{align}
   \phi(z) = \phi_M\coth\!\Big(-\frac{\mu_M}{\sqrt2}\frac{k_M}{|k_M|}\,z + \coth^{-1}k_M\Big)\>,
\end{align}
respectively
\begin{align}
\label{eq:S1MM}
   \phi(z) = \phi_M\frac{k_M}{|k_M|}\coth\!\Big(-\frac{\mu_M z}{\sqrt2} + \coth^{-1}|k_M|\Big)\>.
\end{align}

\subsection{Boundary Conditions}

Using the boundary conditions
\begin{align}
   \phi(z)\Big|_{z=0_-} = \phi(z)\Big|_{z=0_+}\>,
\end{align}
gives the condition
\begin{align}
   \phi_Mk_M = \phi_Vk_V\>,
\end{align}
together with Eq.~(\ref{eq:phimin}) this leads to
\begin{align}
\label{eq:1MBC1}
   \mu_Mk_M = \mu_Vk_V\>.
\end{align}
The second boundary condition 
\begin{align}
   \frac{d\phi}{dz}\bigg|_{z=0_-} = \frac{d\phi}{dz}\bigg|_{z=0_+}\>,
\end{align}
together with Eq.~(\ref{eq:1MBC1}) provides the relation 
\begin{align}
\label{eq:1MBC2}
  -\frac{\mu_M}{|k_M|}\left(1 - k_M^2\right) = \frac{\mu_V}{|k_V|}\left(1 - k_V^2\right).
\end{align}
After some elementary transformations Eqs.~(\ref{eq:1MBC1}) and (\ref{eq:1MBC2}) provide the parameters $k_V$ and $k_M$ characterizing any field solution in terms of the fundamental symmetron parameters and environmental mass densities via $\mu_V$ and $\mu_M$
\begin{align}
\begin{split}
  k_V &= \pm\frac{1}{\sqrt2}\,\sqrt{1 + \frac{\mu_M^2}{\mu_V^2}}\>, \\
  k_M &= \pm\frac{1}{\sqrt2}\,\sqrt{1 + \frac{\mu_V^2}{\mu_M^2}}\>, 
\end{split}
\end{align}
where due to Eq.~(\ref{eq:1MBC1}) 
\begin{align}
  \text{sgn}(k_V) = \text{sgn}(k_M)\>.
\end{align}

\subsection{Final Solution}

Summarizing the findings of the last subsections we finally obtain the field solution
\begin{align}\label{FS1M}
   \phi(z) &= \Theta(+z)\,\varepsilon\,\frac{\mu_V}{\sqrt\lambda}\,\tanh\!\Big(\frac{\mu_V z}{\sqrt2} + \tanh^{-1}|k_V|\Big)  \nonumber\\
   &\quad+ \Theta(-z)\,\varepsilon\,\frac{\mu_M}{\sqrt\lambda}\,\coth\!\Big(\frac{\mu_M|z|}{\sqrt2} + \coth^{-1}|k_M|\Big)\>,
\end{align}
where $\varepsilon$ is defined as 
\begin{align}
  \varepsilon = \text{sgn}(k_V) = \text{sgn}(k_M)\>,
\end{align}
and
\begin{align}
\begin{split}
  |k_V| &= \frac{1}{\sqrt2}\,\sqrt{1 + \frac{\mu_M^2}{\mu_V^2}}\>, \\
  |k_M| &= \frac{1}{\sqrt2}\,\sqrt{1 + \frac{\mu_V^2}{\mu_M^2}}\>.
\end{split}
\end{align}
For comparison with the solutions in the 2-mirror case to be discussed in the following sections, we express Eq~(\ref{FS1M}) also in terms of 
Jacobi elliptic functions 
\begin{align}
   \phi(z) &= \Theta(+z)\,\varepsilon\,\frac{\mu_V}{\sqrt\lambda}\,\textrm{sn}\Big(\frac{\mu_V z}{\sqrt2} + \textrm{sn}^{-1}(|k_V|,1),1\Big)  \nonumber\\
   &\quad+ \Theta(-z)\,\varepsilon\,\frac{\mu_M}{\sqrt\lambda}\,\textrm{ns}\Big(\frac{\mu_M|z|}{\sqrt2} + \textrm{ns}^{-1}(|k_M|,1),1\Big)\>.
\end{align}

\section{2 Mirrors: Symmetric Solution}\label{sec:3}

In this section, we consider symmetric solutions, i.e. obeying $\phi(-z) = \phi(z)$, which arise for two parallel infinitely thick mirrors separated at distance $2d$ in $z$-direction, with $z=0$ being the center between the two mirrors.

\subsection{Vacuum \& Mirror Solutions}

In \cite{Brax:2017hna}, the solution for the vacuum region of low density $\rho_V$ between the mirrors was found to be
\begin{align}
   \phi(z) = \phi_Vk_V\,\textrm{cd}\bigg\{\mu_V\sqrt{1 - k_V^2/2}\,z,\frac{|k_V|/\sqrt2}{\sqrt{1 - k_V^2/2}}\bigg\}\>,
\end{align}
where $k_V$ gives the ratio between the value of the field at the midpoint $z=0$ between the two plates and the vacuum value $\phi_V$. 
Therefore, its absolute value must obey $|k_V|\leq1$.

The solutions within the mirrors can directly be obtained from the corresponding solution in the 1-mirror case Eq.~(\ref{eq:S1MM}) as
\begin{align}
   \phi(z) = \phi_M\frac{k_M}{|k_M|}\coth\!\Big(\frac{\mu_M}{\sqrt2}\left(|z| - d\right) + \coth^{-1}|k_M|\Big)\>.
\end{align}

\subsection{Boundary Conditions}

Using the boundary condition at the mirror surface
\begin{align}
\label{eq:BC2MS1}
   \phi(z)\Big|_{z=d_-} = \phi(z)\Big|_{z=d_+}\>,
\end{align}
provides the condition
\begin{align}
\label{eq:BC2MS1b}
   \phi_Vk_V\,\textrm{cd}\bigg\{\mu_V\sqrt{1 - k_V^2/2}\,d,\frac{|k_V|/\sqrt2}{\sqrt{1 - k_V^2/2}}\bigg\} = \phi_Mk_M\>.
\end{align}
Furthermore, the boundary condition 
\begin{align}
\label{eq:BC2MS2}
   \frac{d\phi}{dz}\bigg|_{z=d_-} = \frac{d\phi}{dz}\bigg|_{z=d_+}\>,
\end{align}
has to be satisfied. Using 
\begin{align}
   \frac{d}{dz}\,\textrm{cd}(z,k) = -\left(1 - k^2\right)\frac{\textrm{sn}(z,k)}{\textrm{dn}^2(z,k)}\>,
\end{align}
and Eq.~(\ref{eq:phimin}) leads after some transformations to the relation
\begin{align}
\label{eq:BC2MS3}
	&-\sqrt2\left(1 - k_V^2\right)|k_V|\,\text{sgn}\big(\textrm{cd}(\alpha,\beta)\big)\,\textrm{sn}(\alpha,\beta) = \nonumber\\
	&\qquad\sqrt{1 - k_V^2/2}\left(\frac{\mu_M^2}{\mu_V^2}\,\textrm{dn}^2(\alpha,\beta) - k_V^2\,\textrm{cn}^2(\alpha,\beta)\right)\>,
\end{align}
where $\alpha = \mu_V\sqrt{1 - k_V^2/2}\,d$ and $\beta = |k_V|/\sqrt{2 - k_V^2}$. The solutions of Eq.~(\ref{eq:BC2MS3}) provide  
possible values of $k_V$. Subsequently, for given $k_V$ the possible values of $k_M$ are obtained by 
\begin{align}
\label{eq:BC2MS4}
   k_M = \frac{\mu_V}{\mu_M}\,k_V\,\textrm{cd}(\alpha,\beta)\>,
\end{align}
where we have employed Eqs.~(\ref{eq:phimin}) and (\ref{eq:BC2MS1b}). 

The discrete set of paired values $(k_V, k_M)$ obtained in this way provides all solutions satisfying the boundary conditions Eqs.~(\ref{eq:BC2MS1}) and (\ref{eq:BC2MS2}).

\subsection{Final Solution}

Summarizing the findings of the previous subsections we finally obtain the field solution
\begin{align}
\label{eq:FS2MS}
   \phi(z) &= \Theta(d - |z|)\,\frac{k_V\mu_V}{\sqrt\lambda}\,\textrm{cd}(\alpha\,z/d, \beta) + \Theta(|z| - d) \nonumber\\
   &\quad\times\frac{k_M}{|k_M|}\frac{\mu_M}{\sqrt\lambda}\,\coth\!\Big(\frac{\mu_M}{\sqrt2}\left(|z| - d\right) + \coth^{-1}|k_M|\Big)\>,
\end{align}
with $\alpha = \mu_V\sqrt{1 - k_V^2/2}\,d$, $\beta = |k_V|/\sqrt{2 - k_V^2}$ and where the possible values of $k_V$ satisfying the boundary conditions 
are given by the solution of
\begin{align}
\label{eq:FS2MSBC1}
	&-\sqrt2\left(1 - k_V^2\right)|k_V|\,\text{sgn}\big(\textrm{cd}(\alpha,\beta)\big)\,\textrm{sn}(\alpha,\beta) = \nonumber\\
	&\qquad\sqrt{1 - k_V^2/2}\left(\frac{\mu_M^2}{\mu_V^2}\,\textrm{dn}^2(\alpha,\beta) - k_V^2\,\textrm{cn}^2(\alpha,\beta)\right).
\end{align}
Subsequently, for given $k_V$ the possible values of $k_M$ are given by Eq.~(\ref{eq:BC2MS4})
\begin{align}
\label{eq:FS2MSBC2}
   k_M = \frac{\mu_V}{\mu_M}\,k_V\,\textrm{cd}(\alpha,\beta)\>.
\end{align}

\section{2 Mirrors: Anti-Symmetric Solution}\label{sec:4}

In this section, we derive anti-symmetric solutions, i.e. obeying $\phi(-z) = -\phi(z)$, for two parallel infinitely thick mirrors separated at distance $2d$ in $z$-direction, with $z=0$ being the center between the two mirrors. 

\subsection{Vacuum \& Mirror Solutions}

In \cite{Brax:2017hna}, the solution for the vacuum region of low density $\rho_V$ between the mirrors was found to be
\begin{align}
   \phi(z) = \pm\frac{\tilde\mu^{(-)}}{\sqrt\lambda}\,\textrm{sn}\bigg\{\frac{\tilde\mu^{(+)}}{\sqrt2}\,z,\frac{\tilde\mu^{(-)}}{\tilde\mu^{(+)}}\bigg\}\>,
\end{align}
where $\tilde\mu^{(\pm)}:=\sqrt{\mu_V^2 \pm \sqrt{\mu_V^4 - 2\lambda\phi_0'^2}}$ and $\displaystyle\phi_0' := \frac{d\phi}{dz}\Big|_{z=0}$. 

Again, the solution inside the mirrors can be obtained directly from the corresponding solution in the 1-mirror case Eq.~(\ref{eq:S1MM}) as
\begin{align}
   \phi(z) = \phi_M\frac{k_M}{|k_M|}\frac{z}{|z|}\coth\!\Big(\frac{\mu_M}{\sqrt2}\left(|z| - d\right) + \coth^{-1}|k_M|\Big)\>.
\end{align}

\subsection{Boundary Conditions}

Using the boundary condition at the mirror surface
\begin{align}
   \phi(z)\Big|_{z=d_-} = \phi(z)\Big|_{z=d_+}\>,
\end{align}
provides the condition
\begin{align}
\label{eq:BC2MA1b}
   \pm\tilde\mu^{(-)}\,\textrm{sn}(\gamma,\delta) = \mu_Mk_M\>,
\end{align}
where $\gamma = \tilde\mu^{(+)}d/\sqrt2$ and $\delta = \tilde\mu^{(-)}/\tilde\mu^{(+)}$ and we have used Eq.~(\ref{eq:phimin}).
Furthermore, also the boundary condition 
\begin{align}
   \frac{d\phi}{dz}\bigg|_{z=d_-} = \frac{d\phi}{dz}\bigg|_{z=d_+}\>,
\end{align}
has to be satisfied. Using 
\begin{align}
   \frac{d}{dz}\,\textrm{sn}(z,k) = \textrm{cn}(z,k)\,\textrm{dn}(z,k)\>,
\end{align}
we obtain the relation
\begin{align}
\label{eq:BC2MA2b}
   \pm\frac{\tilde\mu^{(+)}\tilde\mu^{(-)}}{\sqrt\lambda}\,\textrm{cn}(\gamma,\delta)\,\textrm{dn}(\gamma,\delta) = \phi_M\mu_M\,\frac{k_M}{|k_M|}\left(1 - k_M^2\right)\>.
\end{align}
Using this together with Eq.~(\ref{eq:BC2MA1b}) leads after some transformations to
\begin{align}
	&\tilde\mu^{(+)}\textrm{cn}(\gamma,\delta)\,\textrm{dn}(\gamma,\delta) =  \nonumber\\
	&\qquad\text{sgn}\big(\textrm{sn}(\gamma,\delta)\big)\left(\frac{\mu_M^2}{\tilde\mu^{(-)}} - \tilde\mu^{(-)}\textrm{sn}^2(\gamma,\delta)\right).
\end{align}
The solutions of this equation provide possible values of $\phi_0'$. 
Subsequently, for given $\phi_0'$ (and therefore given $\tilde\mu^{(\pm)}$) the possible values of $k_M$ are obtained from Eq.~(\ref{eq:BC2MA1b})
\begin{align}
\label{eq:AS2Mk}
   k_M = \pm\frac{\tilde\mu^{(-)}}{\mu_M}\,\textrm{sn}(\gamma,\delta)\>.
\end{align}

\subsection{Final Solution}

Summarizing the findings of the previous subsections we finally obtain the field solution
\begin{align}
\label{eq:FS2MA}
 \phi(z) &= \pm\Theta(d - |z|)\,\frac{\tilde\mu^{(-)}}{\sqrt\lambda}\,\textrm{sn}(\gamma\,z/d,\delta) \nonumber\\
   &\quad\pm\Theta(|z| - d)\,\text{sgn}\big(z\,\textrm{sn}(\gamma,\delta)\big) \nonumber\\
   &\qquad\times\frac{\mu_M}{\sqrt\lambda}\,\coth\!\Big(\frac{\mu_M}{\sqrt2}\left(|z| - d\right) + \coth^{-1}|k_M|\Big)\>,
\end{align}
where $\gamma = \tilde\mu^{(+)}d/\sqrt2$ and $\delta = \tilde\mu^{(-)}/\tilde\mu^{(+)}$. Furthermore, 
$\tilde\mu^{(\pm)}:=\sqrt{\mu_V^2 \pm \sqrt{\mu_V^4 - 2\lambda\phi_0'^2}}$ and $\displaystyle\phi_0' := \frac{d\phi}{dz}\Big|_{z=0}$. 
The solutions of 
\begin{align}
\label{eq:FS2MABC1}
	&\tilde\mu^{(+)}\textrm{cn}(\gamma,\delta)\,\textrm{dn}(\gamma,\delta) =  \nonumber\\
	&\qquad\text{sgn}\big(\textrm{sn}(\gamma,\delta)\big)\left(\frac{\mu_M^2}{\tilde\mu^{(-)}} - \tilde\mu^{(-)}\textrm{sn}^2(\gamma,\delta)\right),
\end{align}
provide possible values of $\phi_0'$, while for given $\phi_0'$ (and therefore given $\tilde\mu^{(\pm)}$) the possible values of $k_M$ are given by Eq.~(\ref{eq:AS2Mk})
\begin{align}
\label{eq:FS2MABC2}
   k_M = \pm\frac{\tilde\mu^{(-)}}{\mu_M}\,\textrm{sn}(\gamma,\delta)\>.
\end{align}
For a further discussion concerning the completeness of the 2 mirrors solutions we refer to the corresponding discussion in the previous publication \cite{Brax:2017hna}, which is valid for the solutions obtained herein as well.

\section{A Particular Example}\label{sec:5}

\begin{figure}
\begin{center}
\epsfxsize=8.5 cm \epsfysize=5.2 cm {\epsfbox{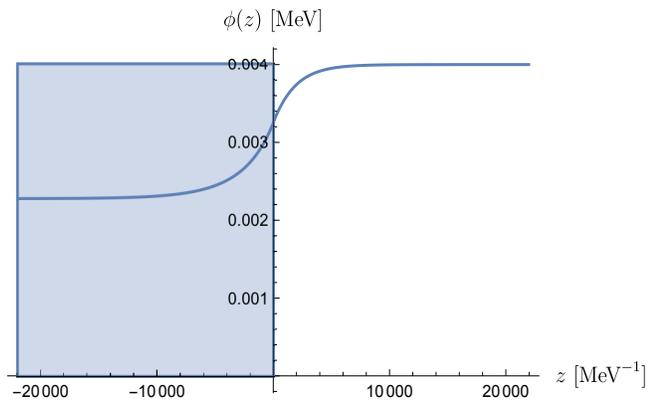}}
\end{center}
\caption{The 1 mirror solution is depicted for the parameters of Tab.~\ref{table:APE}. One can clearly see that the solution approaches a non-vanishing minimum value inside the mirror, which is lower than the corresponding minimum in vacuum. Furthermore, the necessary condition $\phi/M\ll1$ from Eq.~(\ref{eq:WRF}) is indeed satisfied.}
\label{Fig4}
\end{figure}

For illustrative purposes we provide in this section the complete set of solutions for 2 mirrors with the particular parameters as given in Tab.~\ref{table:APE}. The corresponding 1 mirror solution with $k = 0.21$ is depicted in Fig.~\ref{Fig4}.
\begin{table}[!ht]  
\centering
\addtolength{\tabcolsep}{2pt}
\renewcommand{\arraystretch}{1.5}
\begin{tabular}{|c||c|c|c|c|}
  \hline
  Mode & $E$ [MeV$^3$]  & $k_V$ & $k_M$ & $|\phi_0'|$ [MeV$^2$] \\
  \hline\hline
$0^+$ & $3.94\times10^{-10}$  & $\phantom{-}1.00$ & $\phantom{-}1.43$ &    \\
  \hline
$3^-$ & $1.84\times10^{-8}$  & & $-0.14$ & $1.12\times10^{-6}$ \\
  \hline
$4^+$ & $2.41\times10^{-8}$  & $\phantom{-}0.84$ & $\phantom{-}1.36$ &    \\
  \hline
$5^-$ & $2.90\times10^{-8}$  & & $\phantom{-}1.19$ & $9.65\times10^{-7}$  \\
  \hline
$6^+$ & $3.21\times10^{-8}$  & $-0.47$ & $\phantom{-}0.82$ &    \\
  \hline
$7^-$ & $3.26\times10^{-8}$  & & $-0.20$ & $3.96\times10^{-7}$ \\
  \hline
\end{tabular}
\caption{Values of energy $E$ as defined in Eq.~(\ref{eq:RE}) are given for the three solutions for $\rho_\text{eff} = 1.082\times10^{-5}$ MeV$^4$, $M=10$ MeV, $\mu=4\times10^{-4}$ MeV, $\lambda = 10^{-2}$ and mirror distance $d=10^{-8}$ m.}
\label{table:APE}
\end{table}

The "1-dimensional energies" are defined by
\begin{align}
\label{eq:RE}
  E = \int_{-\infty}^\infty \big(\mathcal H(z) - \mathcal H_0(z)\big)\,dz\>,
\end{align}
where the Hamiltonian $\mathcal H(z)$ is given by Eq.~(\ref{Ham}) and
\begin{align}
  \mathcal H_0(z) = - \Theta(|z| - d)\,\frac{\mu_M^4}{4\lambda} - \Theta(d - |z|)\,\frac{\mu_V^4}{4\lambda}\>,
\end{align}
"renormalizes" the energy Eq.~(\ref{eq:RE}) to a finite value. Hereby, we have used that $\mu_i^4/4\lambda$ is the Hamiltonian for the ground state solution in a medium filled with density $\rho_i$.

\begin{figure}
\begin{center}
\epsfxsize=8.5 cm \epsfysize=5.2 cm {\epsfbox{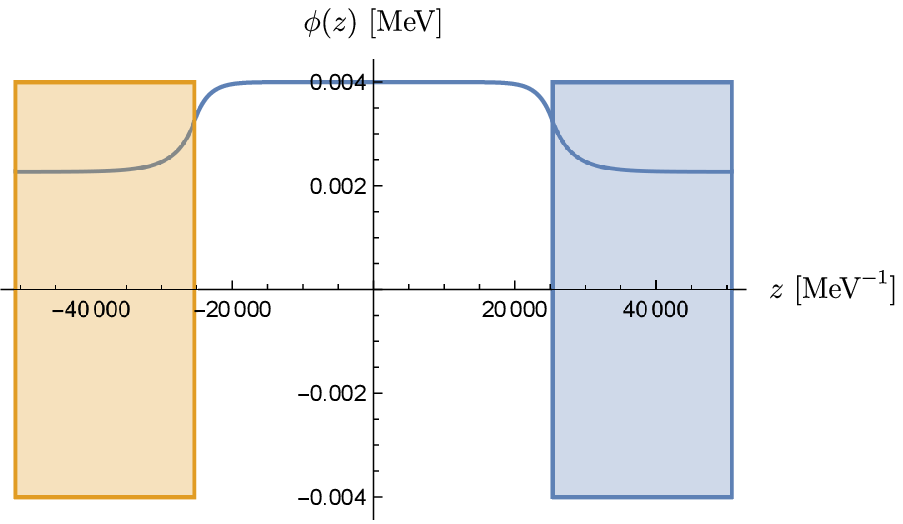}}
\epsfxsize=8.5 cm \epsfysize=5.2 cm {\epsfbox{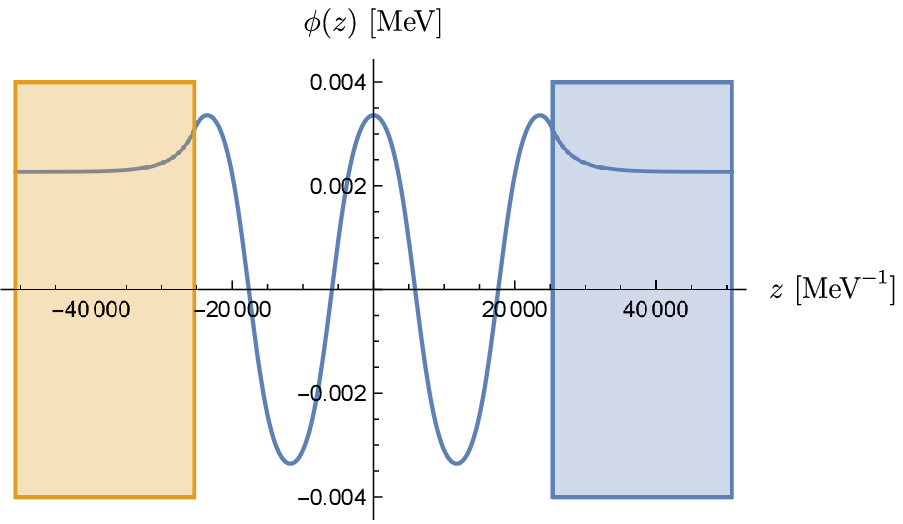}}
\epsfxsize=8.5 cm \epsfysize=5.2 cm {\epsfbox{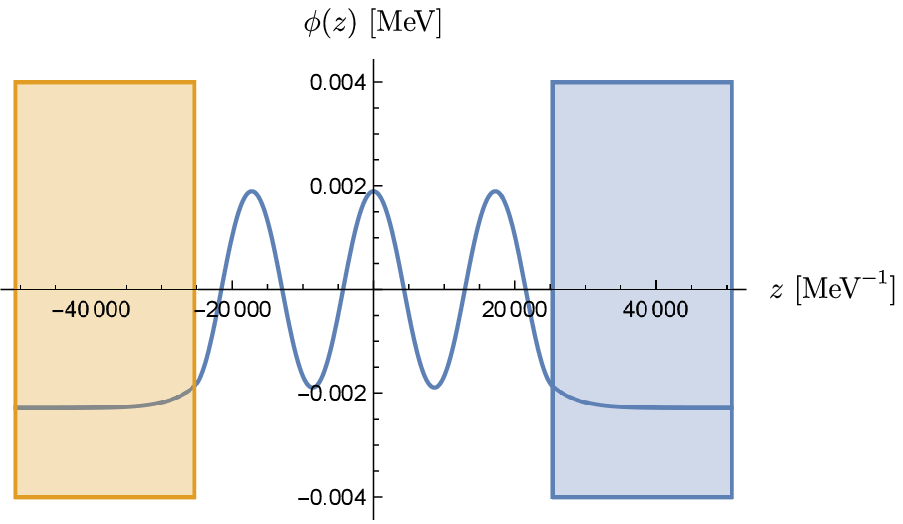}}
\end{center}
\caption{The field profiles for the three symmetric solutions from Tab.~\ref{table:APE} are depicted. Again, the necessary condition $\phi/M\ll1$ from Eq.~(\ref{eq:WRF}) is indeed satisfied.}
\label{Fig5}
\end{figure}

\begin{figure}
\begin{center}
\epsfxsize=8.5 cm \epsfysize=5.2 cm {\epsfbox{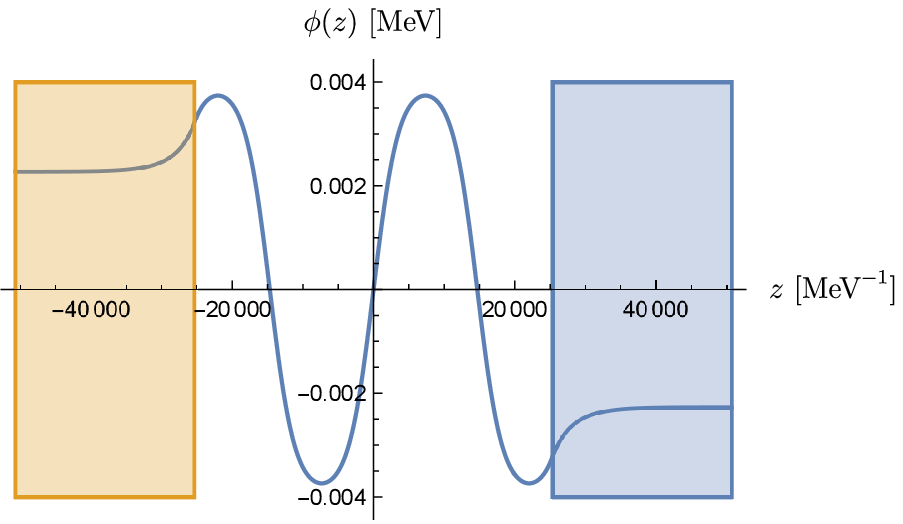}}
\epsfxsize=8.5 cm \epsfysize=5.2 cm {\epsfbox{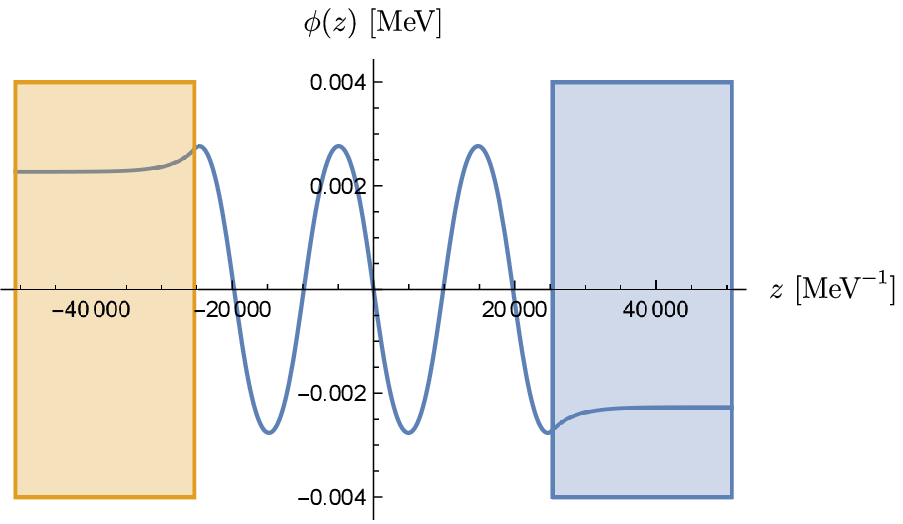}}
\epsfxsize=8.5 cm \epsfysize=5.2 cm {\epsfbox{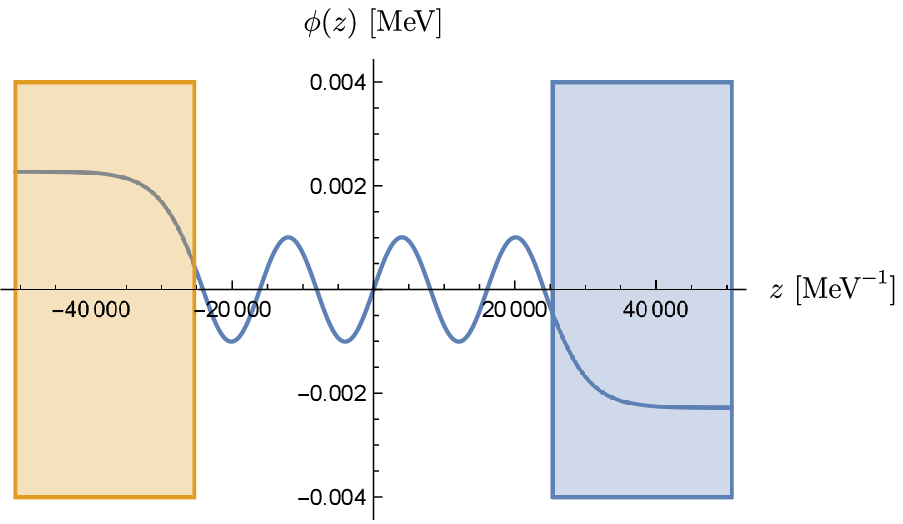}}
\end{center}
\caption{The field profiles for the three anti-symmetric solutions from Tab.~\ref{table:APE} are depicted. Once more, the necessary condition $\phi/M\ll1$ from Eq.~(\ref{eq:WRF}) is indeed satisfied.}
\label{Fig5b}
\end{figure}

The complete set of six solutions is given in Tab.~\ref{table:APE} and the corresponding field profiles are depicted in Figs.~\ref{Fig5} and \ref{Fig5b}. 
These six solutions exhaust the spectrum of possible solutions completely. Each solution is denoted by its number of nodes $0,3,4,5,6,7$ and an upper index $\pm$ to denote its (anti-)symmetry. In Tab.~\ref{table:APE} the solutions are ordered with increasing number of nodes, which corresponds to increasing energy. 
This is in agreement with the results in \cite{Brax:2017hna} where the symmetron is in its symmetric phase inside the mirrors.
Surprisingly, the boundary conditions for the chosen parameter values neither permit a $1^-$ nor a $2^+$ solution.  
One expects that depending on the parameter values and the distance between the mirrors solutions with an arbitrary number of nodes are possible.

\section{Symmetron induced Frequency Shift in \textit{q}BOUNCE}\label{sec:6}

In this section we analyze observable effects of symmetrons in the \textit{q}BOUNCE experiment using the exact solutions obtained herein. This analysis follows the one performed in \cite{Brax:2017hna}. Therefore, the representation will be rather succinct and we refer to this article for further details. 
This analysis can also be used for the GRANIT experiment, which suggests to use magnetic gradient fields for the induction of resonant transitions \cite{Roulier:2014toa}.

In the \textit{q}BOUNCE experiment \cite{Abele:2009dw, Jenke:2011zz, Jenke:2014yel} ultra-cold neutrons fall freely in Earth's gravitational field and are totally reflected by a neutron mirror (this has been reported for the first time in \cite{Nesvizhevsky:2002ef}). The resulting eigenstates are discrete and non-equidistant allowing to apply resonance spectroscopy (for a description of the basic setup we refer to \cite{Jenke:2011zz}). In its version as Rabi-spectroscopy an energy resolution of 3$\times$10$^{-15}$ peV has been obtained \cite{Cronenberg:2018qxf}. In this particular realization of the experiment, the ultra-cold neutrons are allowed to pass a first region, ideally only in the energy ground state of $E_1 = 1.41$ peV, i.e. this region acts as a state selector for the ground state. Subsequently, mechanical vibrations of a glass mirror with a tunable frequency $\omega$ induce Rabi transitions between the ground and a selected excited energy state. Finally, the third region acts again as a state selector for the ground state. Any potential acting on the neutrons, in addition to the gravitational one, induces shifts in the energy levels and thus affects the observable transmission rate of neutrons passing all three regions. Non-observation of such deviations within experimental sensitivity allows to put constraints on these hypothetical additional potentials.

The neutron in a gravitational potential is described by the Schr\"odinger equation 
\begin{align}
   -\frac{\hbar^2}{2m}\frac{\partial^2\psi_n(z)}{\partial z^2} + mgz\,\psi_n(z)  + \delta V(z)\,\psi_n(z) = E_n\psi_n(z)\>,
\end{align}
where $\delta V(z)$ refers to any potential acting on the neutron in addition to the gravitational one. In order to obtain the symmetron induced frequency shift 
we have to extract $\delta V(z)$ caused by the symmetron.

The semi-classical neutron-symmetron coupling can be extracted from the effective symmetron potential Eq.~(\ref{eq:VEFF}) as
\begin{align}
   V_\text{eff} = \frac{1}{2}\frac{m}{M^2}\,\psi^*\psi\,\phi^2\>.
\end{align}
We note that a semi-classical treatment of the neutron involves some subtleties which are discussed at some length in the Appendix of \cite{Brax:2017hna}. Hence, we refer to this article for further information. Finally, the symmetron induced potential $\delta V(z)$ is found to be
\begin{align}
    \delta V(z) = \frac{1}{2}\frac{m}{M^2}\,\phi^2(z)\>.
\end{align}
To first order this leads to a resonance frequency shift in Rabi transitions between two neutron eigenstates $\psi_m$ and $\psi_n$ (see e.g. \cite{landau1991quantenmechanik} for a textbook treatment of perturbation theory)
\begin{align}
   \delta E_{mn} &\equiv E_m - E_n \nonumber\\
   &= \frac{1}{2}\frac{m}{M^2}\int_{-\infty}^\infty dz\,\Big(\big|\psi_m(z)\big|^2 - \big|\psi_n(z)\big|^2\Big)\,\phi(z)^2\>.
\end{align}
For \textit{q}BOUNCE we consider a single mirror filling $z \leq 0$. Hence, we can employ Eq.~(\ref{FS1M}) and finally obtain for the resonance frequency shift
\begin{align}\label{RFS1}
   \delta E_{mn} &= \frac{1}{2}\frac{m}{M^2}\frac{\mu_V^2}{z_0\lambda}\int_0^\infty dz\,\tanh\!\Big(\frac{\mu_V z}{\sqrt2} + \tanh^{-1}|k_V|\Big)^2 \nonumber\\
   &\quad\times\Bigg\{\frac{\displaystyle\text{Ai}\Big(\frac{z - z_m}{z_0}\Big)^2}{\text{Ai}'\Big(-\displaystyle\frac{z_m}{z_0}\Big)^2} - \frac{\displaystyle\text{Ai}\Big(\frac{z - z_n}{z_0}\Big)^2}{\text{Ai}'\Big(-\displaystyle\frac{z_n}{z_0}\Big)^2}\Bigg\}\>,
\end{align}
where $z_0 = \sqrt[3]{\hbar^2/2m^2g}$ and $z_n = E_n/(mg)$. 

We note that the difference in Eq.~(\ref{RFS1}) to the corresponding expression in \cite{Brax:2017hna} amounts to a different shift of the hyperbolic tangent's argument.  
We find that again larger values of $\lambda$ give smaller energy shifts for a given  $\mu_V$. Similarly, increasing $M$ leads to smaller deviations. In this way large regions of the symmetron parameter space can be probed with the \textit{q}BOUNCE experiment. 

In the analysis given so far only the interaction of the neutron with the background field as generated only by the mirrors has been considered. Thereby, the neutron is treated as a probe which does not influence the symmetron field as a source. A proper treatment taking the neutron as a source as well its quantum nature into account constitutes a far more sophisticated problem, which is beyond the scope of this paper. Clearly, more work needs to be done to obtain completely rigorous symmetron bounds using the \textit{q}BOUNCE experiment. As an approximation we treat the neutron as a classical sphere \cite{Brax:2017hna} with diameter $R$, which is a commonly used treatment for neutrons. Such a treatment permits the extraction of a \textit{screening charge} $\mathfrak Q$, given in Eq.~(\ref{eq:SCRC}) and derived in Appendix~\ref{sec:B}. In order to account for the neutron acting as a source for the symmetron one has to replace Eq.~(\ref{RFS1}) as follows
\begin{align}
  \delta E_{mn} \to \mathfrak Q\,\delta E_{mn},\label{eq:ScreenedDeltaE}
\end{align}
for the extraction of the experimental limits. 

\section{Symmetron induced Pressure in CASIMIR experiments}\label{sec:7}

Here, we consider limits that can be obtained by the Casimir And Non-Newtonian force EXperiment (\textsc{Cannex}) \cite{Sedmik:2018kqt} (see also \cite{Klimchitskaya:2019fsm, Klimchitskaya:2019nzu}).
This experiment consists of two parallel plates in a vacuum chamber and has been devised to measure the Casimir force and hypothetical fifth forces.
A symmetron field would induce a pressure between those plates, which can be measured with high precision. 

We approximate the setup in 1 dimension along the $z$-axis as follows. Between the upper surface of the fixed lower mirror at $z = 0$ and the lower surface of the movable upper mirror at $z = a$ prevails vacuum. Above that follows the upper mirror with thickness $D$ and above that at $z > a + D$ vacuum prevails again. 
In order to obtain the induced pressure for the movable upper mirror, we apply the symmetron induced force on a point particle Eq.~(\ref{eq:SFP}), which is derived in Appendix \ref{sec:A}, to the macroscopic mirror as follows
\begin{align}
  \vec f_\phi = -\frac{\rho_M}{M^2}\int_{-\infty}^\infty dx\int_{-\infty}^\infty dy\int_d^{d + D} dz\,\phi\,\partial_z\phi\,\vec e_z\>,
\end{align}
where $\rho_M$ is the mirror density. Consequently, the pressure in $z$-direction is given by
\begin{align}
  p_z = -\frac{\rho_M}{M^2}\int_d^{d + D} dz\,\phi\,\partial_z\phi\>.
\end{align}
The corresponding integral is a surface term and hence trivially carried out with the final result 
\begin{align}
\label{eq:CANP}
  p_z = \frac{\rho_M}{2M^2}\left(\phi^2(d) - \phi^2(d + D)\right).
\end{align}
This agrees for a static field configuration of $\phi$ with Eq.~(\ref{eq:KMTEF}), which reduces in this case to
\begin{align}
  \partial^z T_{zz} = -\partial_z p_z  = \partial_z\ln A\,\rho_M\>.
\end{align}
Using this with Eq.~(\ref{eq:WRF}) indeed agrees with Eq.~(\ref{eq:CANP}).
For $\phi(d)$ we employ the value at the mirror surface of the corresponding two mirror solutions given in Eqs.~(\ref{eq:FS2MS}) and (\ref{eq:FS2MA})
\begin{align}
  \phi^2(d) = \frac{\mu_M^2k_M^2}{\lambda}\>,
\end{align}
where $k_M$ is obtained by Eqs.~(\ref{eq:FS2MSBC1}) and (\ref{eq:FS2MSBC2}) for the symmetric solutions and by Eqs.~(\ref{eq:FS2MABC1}) and (\ref{eq:FS2MABC2}) for the anti-symmetric ones.
For $\phi(d + D)$ we can use instead the value at the mirror surface of the one mirror solution given in Eq.~(\ref{FS1M}) 
\begin{align}
  \phi^2(d + D) = \frac{1}{2\lambda}\left(\mu_V^2 + \mu_M^2\right).
\end{align}
Finally, we obtain for the pressure
\begin{align}
\label{eq:SPII}
  p_z = \frac{\rho_M}{4\lambda M^2}\Big(\mu_M^2\big(2k_M^2 - 1\big) - \mu_V^2\Big)\>.
\end{align}
A numerical analysis has been carried out in parallel to this work \cite{Pitschmann2020} to obtain the exclusion region of the symmetron parameter space by employing the \textsc{Cannex} experiment.

\section{Conclusion}\label{sec:8} 

We have extended a previously performed analysis \cite{Brax:2017hna} for the case of symmetrons in the broken phase in vacuum and in matter. 
We have derived exact analytical solutions to the symmetron field theory in the presence of a one or two mirror system. 
The obtained solutions have been expressed in terms of Jacobian elliptic functions. Again as in \cite{Brax:2017hna} a discrete set of solutions with increasing number of nodes and energies has been found in the two mirror case. Surprisingly, the discrete set of solutions obtained in a particular example does not contain the full discrete spectrum of nodes within finite bounds. 

We have analyzed observable effects induced in the presence of symmetrons for the \textit{q}BOUNCE and  
\textsc{Cannex} experiments. Parallel to this analysis a numerical work has been carried out \cite{Pitschmann2020}, which employs the analytical results obtained herein.

\acknowledgments

We thank Hartmut Abele, Philippe Brax, Christian K\"ading and Ren\'e I. P. Sedmik for fruitful discussions. 

\appendix

\section{Force on a Point Particle in a Symmetron Background}\label{sec:A}

Here, we derive the force on a point particle due to a scalar field, which interacts with matter via the Weyl-rescaling of the metric $g_{\mu\nu}\to \tilde g_{\mu\nu} = A^2(\phi)\,g_{\mu\nu}$ in the matter action. Then, the equations of motion for a point particle follow from
\begin{align}\label{eq:KMTJF}
   \tilde\nabla^\mu\tilde T_{\mu\nu} = 0\>,
\end{align}
in the Jordan frame with metric $\tilde g_{\mu\nu}$. In this frame, the presence of the scalar $\phi$ leads to a change in the space-time geometry. Concerning the matter part of the total action the scalar is absorbed within the re-scaled metric. Hence, a point particle still follows geodesics in space-time but with respect to the re-scaled metric $\tilde g_{\mu\nu}$.

On the other hand, in the Einstein frame we have instead 
\begin{align}\label{eq:KMTEF}
  \nabla^\mu T_{\mu\nu} = \ln A(\phi)_{,\nu}\,T\>,
\end{align}
which can be shown to follow from Eq.~(\ref{eq:KMTJF}) \cite{pitschmann2020RSTG}.
In this frame, the metric determining the space-time geometry is $g_{\mu\nu}$ and consequently the geometry of space-time is unaffected by the scalar, which appears in this case as an explicit degree of freedom and induces a force on the particle, which we will derive next.

The energy-momentum tensor of a point particle propagating along a curve $z^\mu(s)$ is given by
\begin{align}
  T^{\mu\nu} = \frac{m}{\sqrt{-g}}\int ds\,\delta^{(4)}(x - z(s))\,\dot z^\mu\dot z^\nu\>,
\end{align}
where the dot denotes derivation with respect to proper time $s$, i.e. $\displaystyle\frac{d}{ds}$. Using
\begin{align}
  \partial_\nu\frac{1}{\sqrt{-g}} = -\frac{1}{\sqrt{-g}}\,\Gamma_{\nu\lambda}^\lambda\>,
\end{align}
the covariant derivative of the energy-momentum tensor takes the form
\begin{align}
  T^{\mu\nu}_{\phantom{\mu\nu};\nu} &= \frac{m}{\sqrt{-g}}\int ds\,\Big(\dot z^\mu\dot z^\nu\partial_\nu\delta^{(4)}(x - z(s))  \nonumber\\
  &\qquad+\Gamma_{\lambda\nu}^\mu\dot z^\lambda\dot z^\nu\delta^{(4)}(x - z(s))\Big)  \nonumber\\
  &=\frac{m}{\sqrt{-g}}\int ds\,\delta^{(4)}(x - z(s))\left(\ddot z^\mu + \Gamma_{\lambda\nu}^\mu\dot z^\lambda\dot z^\nu\right).
\end{align}
Then, Eq.~(\ref{eq:KMTEF}) reads
\begin{align}
   \frac{m}{\sqrt{-g}}\int ds\,\delta^{(4)}(x - z(s))\left(\ddot z^\mu + \Gamma_{\lambda\nu}^\mu\dot z^\lambda\dot z^\nu - \ln A(\phi)^{,\mu}\right) = 0\>.
\end{align}
This provides the equation of motion
\begin{align}
\label{eq:eom}
  \ddot z^\mu + \Gamma_{\lambda\nu}^\mu\dot z^\lambda\dot z^\nu - \ln A(\phi)^{,\mu} = 0\>.
\end{align}
It is interesting to note that the equation of motion can also be obtained by a variation principle as follows. Since a Weyl-rescaling in the Einstein frame leads to a change of the mass of the point particle $m \to A(\phi)\,m$, the action changes accordingly 
\begin{align}
S_\text{m} = -m\int_a^b ds \to -m\int_a^b A(\phi)\,ds\>.
\end{align}
Variation of the action
\begin{align}
 -m\,\delta\int_a^b A(\phi)\,ds = 0\>,
\end{align}
leads again to the equation of motion Eq.~(\ref{eq:eom}).

Comparison with the four-vector of the force in the Einstein frame
\begin{align}
  f^\mu = m\ddot z^\mu\>,
\end{align}
gives the complete force on a point particle 
\begin{align}
  f^\mu = -m\,\Gamma_{\phantom{\mu}\alpha\beta}^\mu\dot z^\alpha\dot z^\beta + m\ln A(\phi)^{,\mu}\>.
\end{align}
The first term on the right-hand side is the gravitational force with the remaining second term being due to the scalar field $\phi$. 
We would like to note that the acceleration is still proportional to the mass, which means that the weak equivalence principle still holds.

Hence, the force $f_\phi^\mu$ caused by the scalar field alone is given by
\begin{align}
  f_\phi^\mu = m\ln A(\phi)^{,\mu}\>.
\end{align}
In the non-relativistic limit this expression reduces to
\begin{align}\label{eq:scafor}
  \vec f_\phi = -m\vec\nabla\ln A(\phi)\>.
\end{align}
Typically, $A(\phi) \simeq 1$ in which case to leading order $\ln A(\phi) \simeq A(\phi) - 1$. Consequently, we find for the force on a particle caused by a scalar $\phi$ to leading order
\begin{align}
  \vec f_\phi &= -m\vec\nabla A(\phi)\>.
\end{align}
In the case of symmetrons $A(\phi)$ is given by Eq.~(\ref{eq:WRF}) and we finally obtain for the force on a point particle induced by the symmetron field
\begin{align}
\label{eq:SFP}
  \vec f_\phi &= -\frac{m}{M^2}\,\phi\,\vec\nabla\phi\>,
\end{align}
and, respectively, for the corresponding acceleration
\begin{align}
\label{eq:SAP}
  \vec a_\phi &= -\frac{1}{M^2}\,\phi\,\vec\nabla\phi\>.
\end{align}

\section{Symmetron Field of a Neutron}\label{sec:B}

In section \ref{sec:6} we have derived the symmetron induced resonance frequency shift in the \textit{q}BOUNCE experiment. This analysis employed the solutions of symmetron fields as sourced only by the mirrors of the experimental setup. For certain symmetron parameter ranges the neutron acts as a source of the symmetron in a non-negligible way. To take this rigorously into account is beyond the analysis presented in this paper. For an approximative treatment we treat the neutron as a classical sphere. 

The field equations for a static massive sphere with radius $R$ are given by  
\begin{align}
   \frac{d^2\phi}{dr^2} + \frac{2}{r}\frac{d\phi}{dr} = - \mu_i^2\phi + \lambda\,\phi^3\>,
\end{align}
where $\mu_i = \mu_S$ inside the sphere and $\mu_i = \mu_V$ outside. The boundary conditions we take as
\begin{align}
   \phi'(0) &= 0\>, \nonumber\\
   \lim_{r\to\infty}\phi(r) &\to \phi_V\>.
\end{align}
\begin{figure}
\begin{center}
\epsfxsize=8.5 cm \epsfysize=5.2 cm {\epsfbox{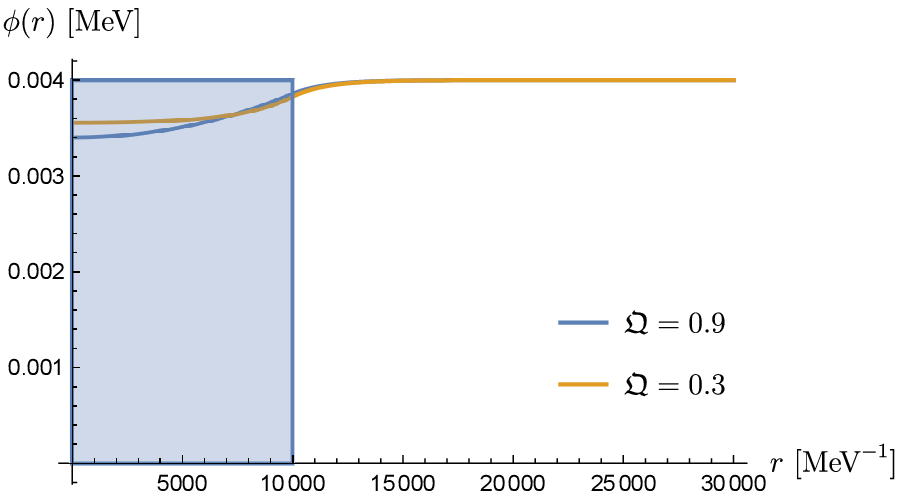}}
\epsfxsize=8.5 cm \epsfysize=5.2 cm {\epsfbox{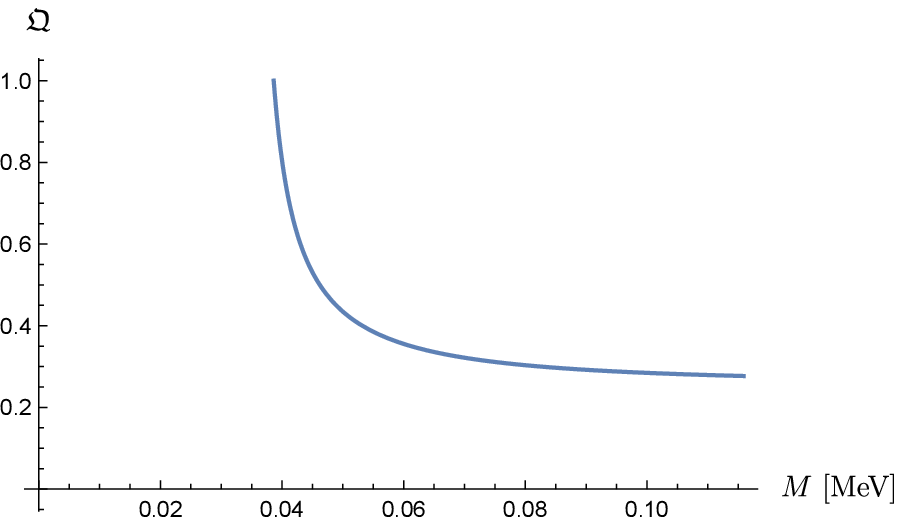}}
\end{center}
\caption{\textit{Top:} The field solution of a sphere $\phi(r)$ is depicted as a function of the radial distance of the center of the sphere for symmetron parameters corresponding to $\mathfrak Q = 0.9$ (blue) and $\mathfrak Q = 0.3$ (yellow). The blue shaded area is bounded by the radius of the sphere $R = 10^{4}$ MeV$^{-1}$ and the vacuum field value $\phi_V$.
\textit{Bottom:} The \textit{screening charge} $\mathfrak Q$ is plotted here as a function of the coupling parameter $M$. For illustrative purposes the parameters taken are $R = 10^{4}$ MeV$^{-1}$, $\rho_S = 2.39\times10^{-10}$ MeV$^{4}$ and $\mu = 4\times10^{-4}$ MeV. Note the lower bound given by $M = \sqrt{\rho_S}/\mu$, below which the symmetron undergoes a phase transition to its symmetric phase.}
\label{FigScr}
\end{figure}
We approximate the effective potential around the minimum value $\mu_i/\sqrt\lambda$ to second order. With  
\begin{align}
   \phi = \varphi + \frac{\mu_i}{\sqrt\lambda}\>,
\end{align}
this corresponds in the equation of motion to the approximation
 \begin{align}
   - \mu_i^2\phi + \lambda\,\phi^3 &= - \mu_i^2\left(\varphi + \frac{\mu_i}{\sqrt\lambda}\right) + \lambda\left(\varphi + \frac{\mu_i}{\sqrt\lambda}\right)^3 \nonumber\\
   &\simeq 2\mu_i^2\varphi\>.
\end{align}
The field equation for a static massive sphere in this approximation is given by
\begin{align}
\label{eomsi}
   \frac{d^2\varphi}{dr^2} + \frac{2}{r}\frac{d\varphi}{dr} = 2\mu_i^2\varphi\>.
\end{align}
With $\varphi = \psi/r$ Eq.~(\ref{eomsi}) reads
\begin{align}
   \frac{d^2\psi}{dr^2} = 2\mu_i^2\psi\>,
\end{align}
with the general solution
\begin{align}
   \phi = A\,\frac{e^{\sqrt2\mu_ir}}{r} + B\,\frac{e^{-\sqrt2\mu_ir}}{r} + \frac{\mu_i}{\sqrt\lambda}\>,
\end{align}
where $A$ and $B$ are arbitrary constants. The solution inside the sphere, which is convergent for $r\to0$ and satisfies the boundary condition $\phi'(0) = 0$ is given by 
\begin{align}
   \phi_<(r) = C\,\frac{\displaystyle\sinh\big(\sqrt2\mu_Sr\big)}{r} + \phi_S\>,
\end{align}
where $C := -2A = 2B$ and $\phi_S = \mu_S/\sqrt\lambda$. The solution outside the sphere, which is convergent for $r\to\infty$ and satisfies $\lim_{r\to\infty}\phi(r) \to \phi_V = \mu_V/\sqrt\lambda$ is given by
\begin{align}
   \phi_>(r) = B\,\frac{e^{-\sqrt2\mu_V r}}{r} + \phi_V\>.
\end{align}
At the surface of the sphere the boundary conditions  
\begin{align}
   \phi_<(R) &= \phi_>(R)\>, \nonumber\\
   \phi'_<(R) &= \phi'_>(R)\>,
\end{align}
must hold, which provides the following expressions for the constants
\begin{align}
   C &= \frac{\phi_V - \phi_S}{\sqrt{2}}\frac{1 + \sqrt2\mu_V R}{\mu_S\cosh\big(\sqrt2\mu_SR\big) + \mu_V\sinh\big(\sqrt2\mu_SR\big)}\>, \nonumber\\
   B &= -\frac{\phi_V - \phi_S}{\sqrt{2}}\frac{\sqrt2\mu_SR\cosh\big(\sqrt2\mu_SR\big) - \sinh\big(\sqrt2\mu_SR\big)}{\mu_S\cosh\big(\sqrt2\mu_SR\big) + \mu_V\sinh\big(\sqrt2\mu_SR\big)} \nonumber\\ &\quad\times e^{\sqrt2\mu_VR}\>.
\end{align}
Consequently, we obtain for the approximative symmetron solution of a sphere 
\begin{widetext}
\begin{align}
\phi(r) = \begin{dcases}
    \phi_S + \frac{\phi_V - \phi_S}{\sqrt{2}}\frac{1 + \sqrt2\mu_V R}{\mu_S\cosh\big(\sqrt2\mu_SR\big) + \mu_V\sinh\big(\sqrt2\mu_SR\big)}\frac{\displaystyle\sinh\big(\sqrt2\mu_Sr\big)}{r}\>,  \qquad \text{for } r\leq R\>, \nonumber\\
    \phi_V - \mathfrak Q\,\frac{\phi_V - \phi_S}{3}\frac{2\mu_S^2R^3}{1 + \sqrt2\mu_V R}\frac{e^{-\sqrt2\mu_V (r - R)}}{r}\>, \qquad \text{for } r\geq R\>.
  \end{dcases}
\end{align}
\end{widetext}
The solution $\phi(r)$ is plotted in Fig.~(\ref{FigScr}) \textit{top} for some representative parameter values. 
Outside the sphere the solution is expressed in terms of the \textit{screening charge}, which is obtained as
\begin{widetext}
\begin{align}
\label{eq:SCRC}
   \mathfrak Q = \frac{3}{2\sqrt2}\frac{1 + \sqrt2\mu_V R}{\mu_S^2R^3}\frac{\sqrt2\mu_SR\cosh\big(\sqrt2\mu_SR\big) - \sinh\big(\sqrt2\mu_SR\big)}{\mu_S\cosh\big(\sqrt2\mu_SR\big) + \mu_V\sinh\big(\sqrt2\mu_SR\big)}\>.
\end{align}
\end{widetext}
The \textit{screening charge} is plotted in Fig.~(\ref{FigScr}) \textit{bottom} as function of parameter $M$. One should note the lower bound given by $M = \sqrt{\rho_S}/\mu$. Below this limit the symmetron undergoes a phase transition to its symmetric phase, in which case the corresponding field solution derived in \cite{Brax:2017hna} has to be employed instead. 
The \textit{screening charge} has limits
\begin{align}
  \mathfrak Q \to \begin{dcases}
    0\>, & \quad \text{for \textit{screened} bodies with }R\gg 1/\mu_S\>, \nonumber\\
    1\>, & \quad \text{for \textit{unscreened} bodies with }R\ll 1/\mu_S\>.
  \end{dcases}
\end{align}
In Eq.~(\ref{eq:SAP}) the symmetron induced acceleration on a small test body is given. Using this with the field outside a sphere as given above we obtain asymptotically for large $r$ the acceleration on the test body
\begin{align}
   \vec a_\phi = -\mathfrak Q\,\frac{\phi_V}{M^2}\frac{\phi_V - \phi_S}{3}\frac{2\sqrt2\mu_V\mu_S^2R^3}{1 + \sqrt2\mu_V R}\frac{e^{-\sqrt2\mu_Vr}}{r}\frac{\vec r}{r}\>.
\end{align}
This dependence of the acceleration on $\mathfrak Q$ justifies the identification of the latter as a \textit{screening charge}.

\newpage

\bibliographystyle{utcaps}
\bibliography{SymmetronII}

\end{document}